\documentclass[final,1p,times]{elsarticle} % do not change
\usepackage{graphicx} % do not change
\usepackage{amssymb} % do not change
\usepackage{amsthm} % do not change
\usepackage{lineno} % do not change

\newcommand {\pp} {\mbox{$p+p$}~}

\newcommand {\AuAu} {\mbox{$Au+Au$}~}

\newcommand {\rootsNN} {\mbox{$\sqrt{s_{NN}}$}}

\journal{Nuclear Physics A} % do not change
\begin{document} % do not change

\begin{frontmatter} % do not change

%% QM09Author: please enter your  
%% Title, author and address info here; please do not use footnotes

% Your Title - please insert
\title{Do \pp Collisions Flow at RHIC? Understanding One-Particle Distributions, Multiplicity Evolution, and Conservation Laws}

% Principle author, and co-authors - please insert
\author{Zbigniew Chaj\c{e}cki and Mike Lisa}

% Address - please insert
\address{The Ohio State University, 191 West Woodruff Avenue, Columbus, Ohio 43210, USA}
\begin{abstract} % do not change
%% Text of abstract goes here - please insert
Collective, explosive flow in central heavy ion collisions manifests itself in the mass dependence
of $p_T$ distributions and femtoscopic length scales, measured in the soft sector
($p_T\lesssim 1$~GeV/c).  Measured $p_T$ distributions from proton-proton collisions differ significantly
from those from heavy ion collisions.  This has been taken as evidence that p+p collisions generate little
collective flow, a conclusion in line with naive expectations.
  We point out possible hazards of ignoring phase-space restrictions due to conservation laws when comparing
high- and low-multiplicity final states.
  Already in two-particle correlation functions, we see clear signals of such phase-space restrictions in
low-multiplicity collisions at RHIC.  We discuss how these same effects, then, {\it must} appear in the
single particle spectra.
  We argue that the effects of energy and momentum conservation actually dominate
the observed systematics, and that $p+p$ collisions may be much more
similar to heavy ion collisions than generally thought.
\end{abstract} % do not change

\end{frontmatter} % do not change

%% QM09: we keep linenumbers at least for initial version
%\linenumbers % do not change

%% start of main text - please insert. 

\section{Introduction and Motivation}
\label{sec:intro}

Most of the interest in the RHIC program falls naturally on collisions between the heaviest nuclei at the highest energies,
where the likelihood of generating a {\it system}, per se, is believed greatest.  However, it is important to understand the
broader context of these measurements; 
the absolute neccessity for extensive systematics is a generic feature of any heavy ion study~\cite{Nagamiya:1988ge,Tannenbaum:2006ch}.
In particular, 
the evolution of the physics as a function of energy may indicate the existence and location of 
predicted critical point in the Equation of State of QCD~\cite{Ritter:2006zz}; 
the evolution as a function of system size 
(e.g. comparing p+p to Au+Au collisions) 
may reveal the emergence of bulk behaviour from the underlying
structure from hadronic collisions.

It is by now well-established that heavy ion collisions at RHIC energies are dominated by
collective hydro-like flow.  
While the degree to which the flowing medium is 
``perfect''~\cite{Gyulassy:2004zy} remains under study, the strongly-coupled nature
of the color-deconfined system is remarkable.  It allows treatment of the system {\it as}
a system, with thermodynamic quantities.  Further, it promises access to the underlying
Equation of State of QCD, together with transport coefficients like viscosity, sometimes
viewed as a complicating factor, but which are in fact is interesting in itself~\cite{Kovtun:2004de}.
In central collisions, the evidence for collective flow comes from the mass dependence of
transverse momentum ($p_T$) distributions and the $p_T$- and mass-dependences
of femtoscopic length scales.  
These may be compared to hydro calculations, but
are often fit with simple parameterizations such as the blast-wave~\cite{Retiere:2003kf} to estimate the strength of the flow.  

Surprisingly, pion HBT measurements in p+p collisions at RHIC
show an identical flow signal as seen
 in Au+Au collisions~\cite{Chajecki:2005zw}.  Indeed, 
similar systematics appear in several hadron-hadron measurements~\cite{Chajecki:2009zg}!
This appears to be at variance with blast-wave fits
to $p_T$ spectra~\cite{Adams:2003xp}, which suggest a much smaller transverse flow in p+p collisions.  Here, we suggest that
the apparent difference between spectra from p+p and Au+Au collisions may be understood in terms of
energy and momentum conservation effects, which are naturally stronger for the smaller system.
For more details on this study, see~\cite{Chajecki:2008yi}.

%%%%%%%%%%%%%%%%%%%%%%%%%%%%%%%%%%%%%%%%%%%%%%%%%%%%%%%%%%%%%%%%%%%%%%%%%%%%%%%%
\begin{figure}[t!]
\centering
\includegraphics[width=0.65\textwidth]{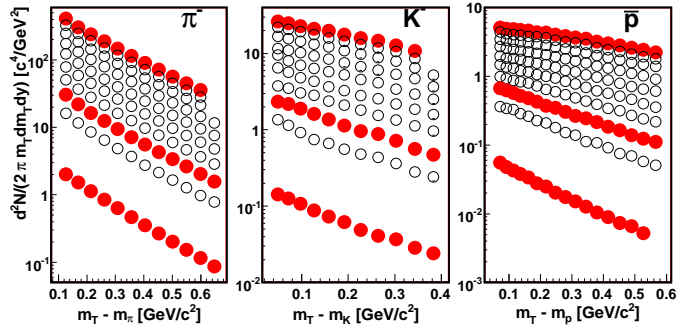}
%\vspace*{-0.3cm}
\caption{Pion (left), kaon (center) and antiproton (right) $m_T$ distributions measured by the STAR
Collaboration for \rootsNN=200~GeV collisions~\cite{Adams:2003xp}.  The lowest datapoints represent
minimum-bias \pp collisions, while the others come from \AuAu collisions of increasing
multiplicity.  
%%Filled datapoints are for the top 5\% and 60-70\% highest-multiplicity \AuAu collisions, and for the \pp collisions.
\label{fig:STARPIDspectra}}
\end{figure}
%%%%%%%%%%%%%%%%%%%%%%%%%%%%%%%%%%%%%%%%%%%%%%%%%%%%%%%%%%%%%%%%%%%%%%%%%%%%%%%%

\section{A fairer comparison of spectra from A+A and p+p collisions}
\label{sec:Compare}

\begin{figure}
\begin{minipage}[t]{0.49\textwidth}
\centering
\includegraphics[width=.95\textwidth]{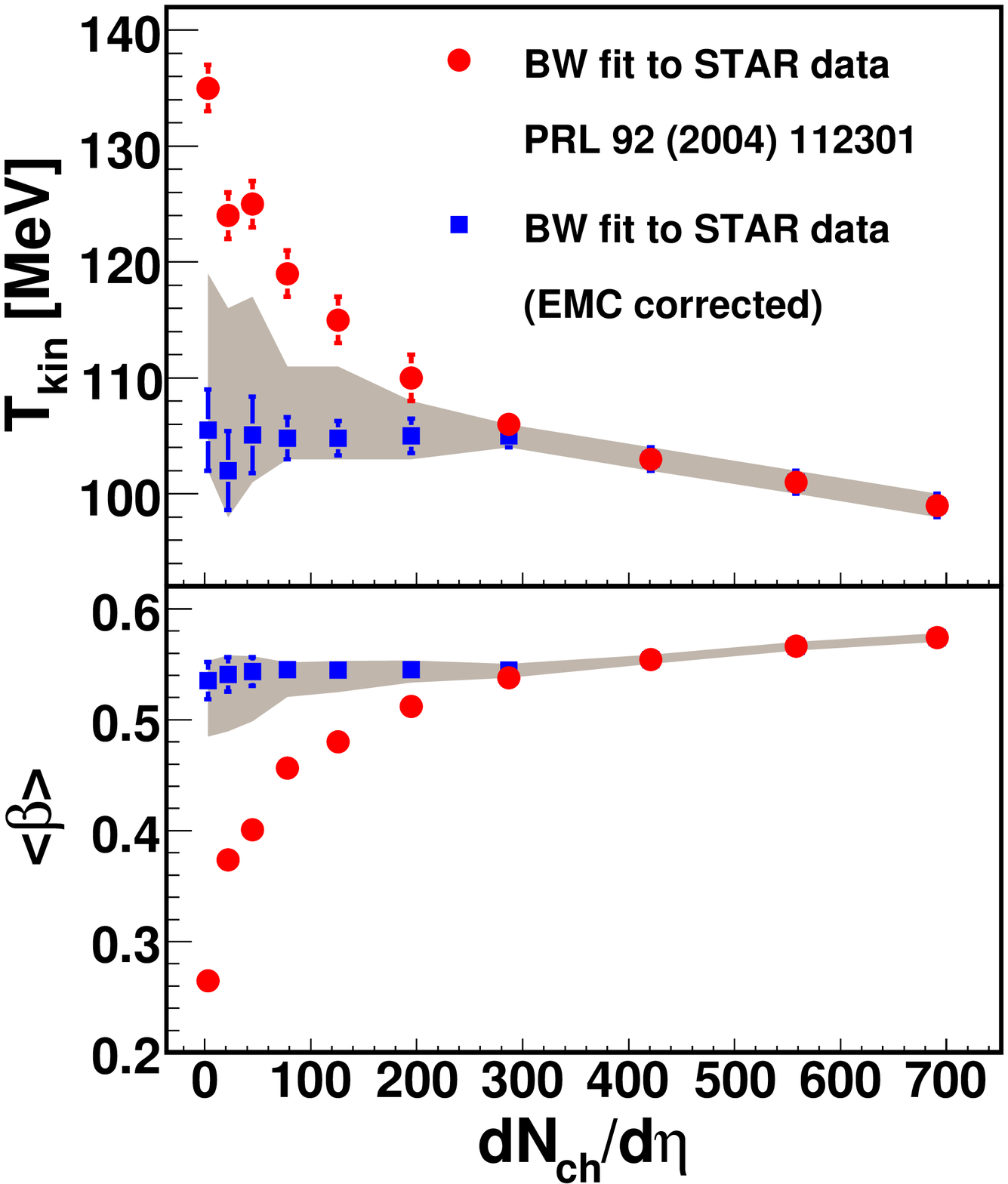}
\vspace*{-0.3cm}
\caption{Circles: temperature (top panel) and flow (bottom panel) parameters of
a blast-wave~\cite{Retiere:2003kf} fit to spectra of Figure~\ref{fig:STARPIDspectra},
as a function of the event multiplicity. Squares: the same fits
to ``corrected'' spectra. Figure from~\cite{Chajecki:2008yi}.
%%%Systematic errors shown by shaded region.
\label{fig:STARbeta}}
\end{minipage}
\hspace{\fill}
\begin{minipage}[t]{0.49\textwidth}
\centering
\includegraphics[width=0.97\textwidth]{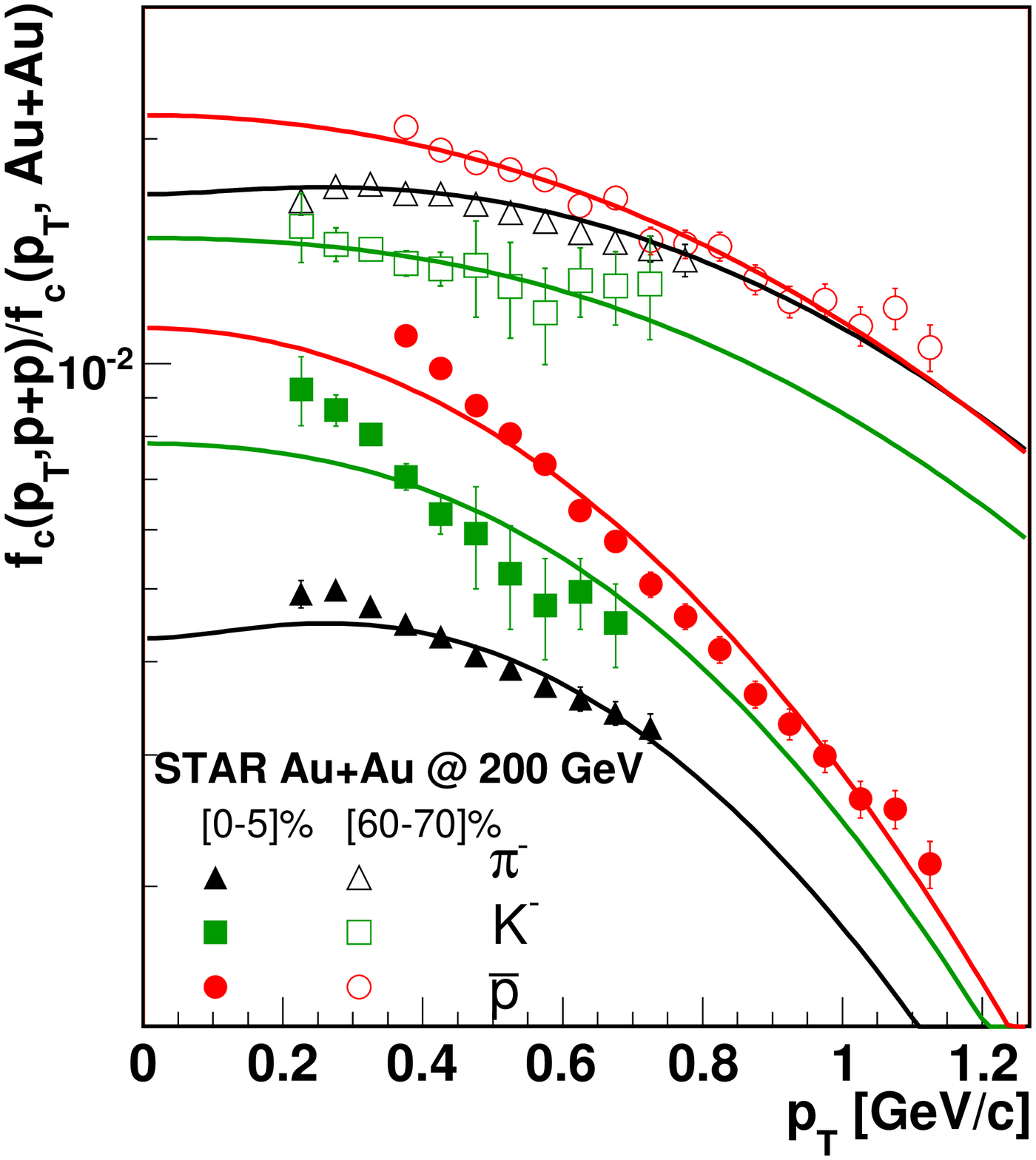}
\vspace*{-0.3cm}
\caption{The ratio of the $p_T$ distribution from minimum-bias \pp collisions to the
distribution from 0-5\% (filled datapoints)
and 60-70\% (open datapoints) highest multiplicity \AuAu collisions.
Figure from~\cite{Chajecki:2008yi}.
%%%c.f. Figure~\ref{fig:STARPIDspectra}.
%The ratio of the kaon spectra from \pp and 0-5\% \AuAu collisions (solid green squares) has been scaled by a factor 1.7 for clarity.
%Curves represent a calculation of this ratio (ratio of EMCIC factors) using Equation~\ref{eq:ratio}.
\label{fig:SoftRatio}}
\end{minipage}
\vspace*{-0.3cm}
\end{figure}

Figure~\ref{fig:STARPIDspectra} shows transverse momentum spectra for pions, kaons and protons measured by the
STAR Collaboration for collisions at $\sqrt{s_{NN}}=200$~GeV.  The spectral shapes evolve as the multiplicity is
increased from p+p collisions (at the bottom) to the highest-multiplicity Au+Au collisions (top).  A blast-wave~\cite{Retiere:2003kf}
fit to these spectra indicates a steadily increasing (decreasing) flow velocity (freezeout temperature) with increasing
multiplicity, as shown by the red circles on Figure~\ref{fig:STARbeta}.  However, these fits entirely neglect effects of
phasespace restrictions due to energy and momentum conservation, whose significance steadily increases with decreasing
multiplicity.

In the approximation that dynamics and kinematic constraints can be factorized, the measured single-particle
distribution $\tilde{f}_c$ from an $N$-particle final state is related to the ``parent'' distribution $\tilde{f}$
according to~\cite{Chajecki:2008yi,Chajecki:2008vg}
\noindent
\begin{equation}
\label{eq:EMCIC2}
\tilde{f}_c\left(p_i\right) \propto \tilde{f}\left(p_i\right) \cdot 
%%%%%\left(\frac{N}{N-1}\right)^2 %%%\cdot
 \exp\left[\frac{-1}{2(N-1)}\left(
\frac{p^2_{i,x}}{\langle p_x^2 \rangle}+\frac{p^2_{i,y}}{\langle p_y^2 \rangle}+\frac{p^2_{i,z}}{\langle p_z^2 \rangle}
+\frac{\left(E_i-\langle E \rangle\right)^2}{\langle E^2 \rangle -\langle E \rangle^2}\right)\right] ,
%%%\end{align}
\end{equation}
where $\langle p^n_\mu\rangle$ are average quantites of energy and 3-momentum.

We now use this formula to test the extreme postulate that the parent distributions-- which reflect the underlying dynamics-- are
{\it identical} for p+p and Au+Au collisions at all centralities.  In this case, the ratio of two {\it measured} spectra $\tilde{f}_{c,1}$
and $\tilde{f}_{c,2}$, from events with multiplicites $N_1$ and $N_{2}$, will be simply the ratio of their phasespace factors:
\begin{eqnarray}
\label{eq:ratio}
\frac{\tilde{f}_{c,1}\left(p_T\right)}{\tilde{f}_{c,2}\left(p_T\right)} \propto 
%%%%    K\times\left(\frac{\left(N_2-1\right)N_1}{\left(N_1-1\right)N_2}\right)^{2} \times % \\ \nonumber
  &\exp\left[
%%%e^{
\left(\frac{1}{2\left(N_2-1\right)}-\frac{1}{2\left(N_1-1\right)}\right) \cdot
 \right.  \\ \nonumber
  &\left. 
\left(\frac{2p_T^2}{\langle p_T^2\rangle}
+ \frac{E^2}{\langle E^2\rangle-\langle E\rangle^2}
- \frac{2E\langle E\rangle}{\langle E^2\rangle-\langle E\rangle^2}
+ \frac{\langle E\rangle^2}{\langle E^2\rangle-\langle E\rangle^2}
\right)
%%%}
\right] 
\quad .
\end{eqnarray}

The datapoints in Figure~\ref{fig:SoftRatio} show  $\pi$, K and p spectra from p+p (full points) and mid-central Au+Au (open points) collisions,
divided by the spectra from the most central Au+Au collisions.  Curves represent Equation~\ref{eq:ratio}, with 
$\langle p_T^2\rangle=0.12~{\rm GeV}^2$,
$\langle E^2\rangle=0.43~{\rm GeV}^2$ and
$\langle E\rangle=0.61~{\rm GeV}$.
According to our postulate, the only difference between the different-multiplicity collisions is, in fact, the multiplicity:
$N_{\rm cent}=500$;
$N_{\rm periph}=18$;
$N_{\rm p+p}=10$.

The curves well describe the shape, magnitude, multiplicity-, and mass-dependence of the changes in the spectra.  This indicates that
the multiplicity evolution of spectral shapes is driven much more by phasespace restrictions due to energy and momentum distributions
than by any real change in dynamics, a rather stunning suggestion.  To emphasize the point, Figure~\ref{fig:STARbeta} shows
blast-wave parameters from fits to the ``corrected'' spectra generated by dividing the measured distributions
by the phase-space factor.

%%%%%%%%%%%%%%%%%%%%%%%%%%%%%%%%%%%%%%%%%%%%%%%%%%%%%%%%%%%%%%%%%%%%%%%%%%%%%%%%
%%%%%%%%%%%%%%%%%%%%%%%%%%%%%%%%%%%%%%%%%%%%%%%%%%%%%%%%%%%%%%%%%%%%%%%%%%%%%%%%

\section{``The system''}

The evolution of the spectral shapes for $p_T\lesssim 1$~GeV/c (at higher $p_T$, other physics takes over~\cite{Chajecki:2008yi})
with event multiplicity is almost perfectly described by Equation~\ref{eq:ratio}, when
the parameters $N$, $\langle p_T^2\rangle$,  $\langle E^2\rangle$ and  $\langle E\rangle$ are adjusted.  The only parameter which
changes with multiplicity cut is $N$; the rest are constant.
But where do these parameters come from?  Are they reasonable?

While it may seem that such parameters may be extracted directly from the data, this is not so.  Firstly, it is important
to include ``primary'' particles in the consideration of phasespace contraints, and to include all particles (including any
photons, neutrinos, etc).  Further, the kinematic quantities $\langle p_T^2\rangle$, etc, are averages over the (unmeasured)
{\it parent} distribution, not the measured $\tilde{f}$.  Finally, there arises the question: ``what is the system for which
a finite amount of energy and momentum is shared between $N$ particles?''  It is likely {\it not} the entire set of particles
emitted from a collision; the ``system'' decaying into the mid-rapidity region is not affected by beam fragmentation.  Rather,
the beam (and, likely, jets) steal some energy away from the smaller system of $N$ particles, which then statistically share
energy.  Thus, the {\it measured} quantities $\langle p_T^2\rangle_c$ serve as a guide to their corresponding parameters in
Equation~\ref{eq:ratio}, but need not fix them.

Further guidance comes from PYTHIA~\cite{Sjostrand:2000wi} simulations, which return quantities within a factor of $\sim 2$
of the ones we use.  The numbers are also consistent with an estimate assuming a Maxwell-Boltzman distribution for the
underlying ``system,'' with temperatures $T=0.15\div 0.35$~GeV.

These issues are discussed in more detail in~\cite{Chajecki:2008yi}.  Thus, while we may use various estimates to
validate our ``reasonable'' parameters, we cannot derive them directly from the measured spectra themselves.  In principle,
they {\it can} be extracted rather directly from two-particle correlation functions.  At this meeting and previously~\cite{Chajecki:2005zw},
STAR has shown very clear phasespace-induced signals in two-pion correlation functions from p+p collisions.  Work is well
underway to extract ``system''
kinematic parameters from the two-particle correlation functions and use them to calculate phasespace effects single-particle spectra.

\section{Summary and discussion}
\label{sec:Summary}

The observed evolution of the $p_T$ distributions measured at RHIC may arise due to changes in the
dynamics as the system varies from p+p to central Au+Au, differences in phasespace constraints, or both.  Most
interpretations, based for example on blast-wave fits to the spectra, assume a dynamical origin for the
spectra differences, but ignore effects of kinematics.  We have shown that phasespace constraints due to energy
and momentum conservation, can {\it alone} explain most of the multiplicity evolution of the spectra.  Any additional
change to the spectra due to dynamical effects must be very small.  This claim will become much more compelling
if one can extract, directly from measured two-pion correlation functions, the system kinematic parameters that
we are now claiming to be ``reasonable.''

We remind the reader that a purely phasespace-based explanation for the spectra evolution
breaks down for $p_T > \sim 1.5$~GeV/c~\cite{Chajecki:2008yi}, where non-bulk physics becomes more dominant.

Since we argue that the parent spectra are, themselves, almost identical, we are not surprised to find that
blast-wave parameters for p+p and Au+Au collisions are almost identical.
The degree to which this implies flow in p+p collisions (or the lack of it in Au+Au collisions) remains unclear.
Since any freezeout scenario should simultaneously describe spectra and femtoscopic measurements,
input from two-pion correlation functions should shed more light on this intriguing question.

%\section*{Acknowledgments} 
%Thank the STAR collaboration for ....

%\bibliographystyle{elsarticle-harv}%natbib}
%\bibliographystyle{elsart-num}%natbib}
%\bibliographystyle{elsart-num-sort}%natbib}
%\bibliographystyle{elsart-num-names}%natbib}
%\bibliographystyle{natbib}
%\bibliography{bibliography}

%% \begin{thebibliography}{00} % do not change 
%% %\bibitem{label} 
%% \end{thebibliography} % do not change 

\end{document}